\documentclass[12pt]{article}
\usepackage[margin=1in]{geometry}

\begin{document}

\begin{center}{\Huge\bf de Sitter symmetry and neutrino oscillations} \end{center}

\begin{center}Felix M Lev\end{center}
\begin{center}Artwork Conversion Software Inc. \\509 N. Sepulveda Blvd
Manhattan Beach CA 90266 USA\end{center} 
\begin{center}Email: felixlev314@gmail.com\end{center}

\begin{abstract}
Although the phenomenon of neutrino oscillations has been confirmed in many experiments,
the theoretical explanation of this phenomenon in the literature is essentially model dependent and is
not based on rigorous physical principles. We propose an approach where the neutrino is
treated as a massless elementary particle in anti-de Sitter (AdS) invariant quantum theory. In contrast to
standard Poincare invariant quantum theory, an AdS analog of the mass squared changes over time
even for elementary particles. Our approach naturally explains why, in contrast to the neutrino,
the electron, muon and tau lepton do not have flavors changing over time and why the number of solar neutrinos reaching the Earth is around a third of the number predicted by the standard solar model.
\end{abstract}

\begin{flushleft} Keywords: quantum de Sitter symmetry; neutrino oscillations; solar neutrino problem\end{flushleft}

\section{Problem statement}
\label{intr}

The phenomenon of neutrino oscillations is discussed in a vast literature and, as noted by
most authors, the theory of this phenomenon contains essential uncertainties
(see e.g., \cite{Greiner,Capozzi}). We first describe the crucial difference between this theory
and standard particle theory. 

The goal of QED, and electroweak theory is to find the S-matrix
describing transitions between different sets of free elementary particles. By definition, 
an elementary particle is described by an irreducible representation (IR) of the Poincare algebra.
If $P_0$ is the energy operator and ${\bf P}$ is the momentum operator then 
$W=P_0^2-{\bf P}^2$ is the Casimir operator of the Poincare algebra, i.e.,
it commutes with all operators defining the representation of the algebra
for the system under consideration. As follows from the Schur lemma, all elements 
in the representation space of an IR are eigenvectors of $W$ with some eigenvalue
$w$. The existence of tachyons (for which $w<0$) is a problem while for known elementary
particles, their mass $m$ is defined as a positive number $m$ such that $m^2=w$.

In standard theory
of weak interactions, the electron, muon and tau neutrinos with different flavors ($\nu_e,\nu_{\mu},\nu_{\tau}$)=($f_1,f_2,f_3$) 
are treated as elementary particles, and the processes of their creation/annihilation
are described by Feynman diagrams involving the $(W,e,\nu_e)$, $(W,\mu,\nu_{\mu})$
and $(W,\tau,\nu_{\tau})$ vertices. In this theory (and in the Standard Model), the electron, muon and tau lepton quantum numbers are conserved physical quantities. However, the phenomenon of neutrino oscillations shows that conservation of those quantum numbers can be only approximate. 

For explaining this phenomenon, a fundamentally new property was introduced into the theory: now neutrinos with definite flavors are treated not as elementary particles but as direct sums of neutrino states ($\nu_1,\nu_2,\nu_3$) such that
$\nu_i$ ($i$=1,2,3) is an elementary particle with a definite mass $m_i$, i.e.,
\begin{equation}
f_i=\sum_{j=1}^3\oplus U_{ij}\nu_j\,\, (i=1,2,3)
\label{directsum}
\end{equation}
where $U_{ij}$ are elements of a complex $3\times 3$ matrix.
The mathematical meaning of this expression is as follows. Each state $\nu_j$ 
 belongs to a Hilbert space $H_j$ of an IR with the mass $m_j$ while $f_i$ belongs
to the direct sum of the spaces $H_j$. 

The principle of superposition in quantum theory does not prohibit states which are direct
sums of IRs. So far, such direct sums have been used only in QCD to describe
mixing of quarks by Cabibbo angle and Cabibbo–Kobayashi–Maskawa matrix.
A crucial difference between such direct sums and direct sums of neutrino IRs is as follows. 
Since quarks cannot be in free states, quarks belonging to the same 
direct sum are inside the same nucleon or meson and the distances between
such quarks cannot exceed the size of the corresponding nucleon or meson.
On the other hand,
there are no theoretical limitations on distances between neutrinos belonging to the
same direct sum of IRs. 

The concept of direct sum of Hilbert spaces crucially differs from the concept of tensor
product of Hilbert spaces which is a standard concept in
quantum theory. In particular, if $H_1$ and $H_2$ are representation spaces
for IRs with the masses $m_1$ and $m_2$ then the tensor product of $H_1$ and $H_2$
describes a system of two elementary particles while the direct sum of $H_1$ and $H_2$
describes one particle which is not elementary because it is a quantum superposition of 
two elementary particles. 

The new treatment of neutrinos implies that in the Feynman diagrams involving the 
$(W,e,\nu_e)$, $(W,\mu,\nu_{\mu})$ and $(W,\tau,\nu_{\tau})$ vertices, the neutrinos are not elementary particles, but nontrivial direct sums defined by Eq. 
(\ref{directsum}). The theory of such diagrams has been proposed in \cite{Naumov} and
here the neutrinos are superpositions of not plane waves but wave packets. As noted
in \cite{Naumov}, one might expect that in this approach several problems will be solved. 

In view of the status of the neutrino theory, the following questions arise:
\begin{itemize}
\item
Why only neutrinos have such an unusual status
while the other leptons in these vertices $(e^{\pm},\mu^{\pm},\tau^{\pm})$
are still treated as elementary particles.
\item
Why, in contrast to all other particle theories, the theory involving neutrinos is formulated not in terms of Feynman diagrams involving the elementary particles ($\nu_1,\nu_2,\nu_3$) but
in terms of ($\nu_e,\nu_{\mu},\nu_{\tau}$) which are now treated not as elementary particles.
\item 
It is not clear whether the numbers $U_{ij}$ should be defined by a new theory or
they can be defined only by fitting experimental data. 
\end{itemize}

Let us stress again that Eq. (\ref{directsum}) does not imply that each neutrino is a
composite state of states ($\nu_1,\nu_2,\nu_3$). This expression shows that each neutrino
can be detected only {\it in one of the states} ($\nu_1,\nu_2,\nu_3$) with the
probabilities defined by the numbers $U_{ij}$. In theories of entanglement, it is explained
that, as follows from the principle of the wave function collapse, a measurement of the
first entangled particle automatically reduces the wave functions of other particles
entangled with the first one, even if they are very far from the first particle. Analogously,
after a neutrino takes part in any interaction, the remaining wave function cannot be
a superposition Eq. (\ref{directsum}) anymore and only one of the wave functions
$\nu_j$ (if any) can survive.

In view of the fact that, in the processes of neutrino oscillations, neutrinos can travel
long distances during a long time, a question arises why, in spite of the phenomenon
of the wave function collapse, those neutrinos remain superpositions of the states ($\nu_1,\nu_2,\nu_3$) during all this time. Even if those neutrinos do not interact with matter,
they necessarily interact with the background gravitational field, and, as follows from
the principle of the wave function collapse, the wave functions of such neutrinos must 
collapse even after very weak interactions.

The above discussion poses a problem whether the phenomenon of neutrino
oscillations can be explained if neutrinos are still treated as elementary particles. We believe
that standard particle theories are not quite natural in view of the following.
Those theories are based on Poincare symmetry, where elementary particles
are described by IRs of the Poincare group or its
Lie algebra. In his famous paper "Missed Opportunities" \cite{Dyson} Dyson notes that de Sitter (dS) and anti-de Sitter (AdS) theories
are more general (fundamental) than Poincare one even from pure mathematical considerations because  
dS and AdS groups are more symmetric than Poincare one. The transition from the former to the latter
is described by a procedure called contraction when a parameter $R$ (see below) goes to infinity. At the same time, since dS and AdS groups are semisimple, they have a maximum possible symmetry and cannot be obtained from more symmetric groups by contraction. 

The paper \cite{Dyson} appeared in 1972 (i.e., more than 50 years ago) and, in view of Dyson's results, a question arises why general theories
of elementary particles are still based on Poincare symmetry and not dS or AdS
ones. Probably, physicists believe that, since $R$ is much greater than even sizes of stars, dS and AdS symmetries can play an important role only in cosmology and there is no need to use them for describing elementary particles. 
We believe that this argument is not consistent because usually more general theories shed a new
light on standard concepts. The discussion in our publications and, in particular, in this paper 
is a good illustration of this point.

In \cite{book} it has been proposed the following 

{\bf Definition:} {\it Let theory A contain a finite nonzero parameter and theory B be obtained from theory A in the formal limit when the parameter goes to zero or infinity. Suppose that, with any desired accuracy, theory A can reproduce any result of theory B by choosing a value of the parameter. On the contrary, when the limit is already
taken, one cannot return to theory A, and theory B cannot reproduce all results of theory A. Then theory A is more general than theory B and theory B is a special degenerate case of theory A}. 

As argued in \cite{book,DS}, in contrast to Dyson's approach based on Lie groups, the approach to symmetry on quantum level should be based on Lie algebras. Then
it has been proved that, on quantum level, dS and AdS symmetries are more general (fundamental) than Poincare symmetry, {\it and this fact has nothing to do with the comparison of dS and AdS spaces with Minkowski space}. It has been also proved that 
classical theory is a special degenerate case of quantum one in the formal limit $\hbar\to 0$ and
nonrelativistic theory is a special degenerate case of relativistic one in the formal limit $c\to\infty$. In the
literature the above facts are explained from physical considerations but, as shown in  \cite{book,DS},
they can be proved mathematically by using properties of Lie algebras.

The goal of the present paper is to investigate whether the phenomenon of neutrino
oscillations can be explained within the AdS quantum theory by treating
neutrinos as elementary particles. In Sec. \ref{dirsum} we describe
problems in treating neutrino oscillations in the framework of current approaches. In Sec. \ref{AdS}
we explain why quantum theory based on de Sitter symmetries is more general and natural
than standard quantum theory. In Secs. \ref{massless} and \ref{matrix} we describe a formalism
for describing the massless neutrino as an elementary particle in AdS invariant quantum theory.
In Sec. \ref{flavors} we discuss how the problem of neutrino flavors is treated in our approach
and explain that the solar neutrino problem has a natural explanation.

\section{Direct sum of Hilbert spaces in neutrino oscillations}
\label{dirsum}

In the present treatments of neutrino oscillations, it is assumed that all the components
of the direct sum in the neutrino wave function have different masses $m_j$. 
The states $\nu_j$ are not characterized by flavor and they cannot 
have different quantum numbers except $m_j$ because in that case their superpositions will be prohibited. 
Since the neutrino flavor quantum numbers are not conserved,
they do not distinguish the states ($\nu_e,\nu_{\mu},\nu_{\tau}$) anymore, and 
those states differ only because they are defined 
by different decompositions in Eq. (\ref{directsum}). 
The phenomenon of neutrino oscillations shows that those states 
do not have definite masses and can transform to each other.  The existing quantum theory does not specify physical criteria which should be used for
constructing the matrix $U$, and currently the numbers $U_{ij}$ are defined only
from fitting the experimental data. In the literature, the matrix $U$ is called 
the Pontecorvo–Maki–Nakagawa–Sakata matrix, PMNS matrix or lepton mixing matrix. It is the analogue of the Cabibbo–Kobayashi–Maskawa matrix which, as noted above, describes mixing of quarks. If the PMNS matrix were the identity matrix, then the flavor eigenstates would be the same as the mass eigenstates. However, experiment shows that they are not. In
\cite{Capozzi} it is discussed in detail which numbers $U_{ij}$ are well defined by experimental data, which of them are defined with uncertainties and which of those numbers
are not known yet.

Since there should be a one-to-one relation between the states ($\nu_e,\nu_{\mu},\nu_{\tau}$) 
and ($\nu_1,\nu_2,\nu_3$), the matrix $U$ should be invertible and then, as follows from Eq. (1), 
\begin{equation}
\nu_i=\sum_{j=1}^3(U^{-1})_{ij}f_j\,\, (i=1,2,3)
\label{inverse}
\end{equation}
There are no physical criteria requiring that all the $\nu_j$ in Eq. (\ref{directsum}) have the same 
momentum. For example, the author of \cite{Kayser} poses the following questions:
''Why should one assume that the different mass eigenstates $\nu_j$ in a beam have
a common momentum but different energies? Why not assume they have a common energy but
different momenta? Or different momenta {\it and} different energies? And what oscillation pattern is predicted if one does make one of these alternate assumptions?'' To this list of questions, one can add the following: ''Why not assume that different mass eigenstates have a common velocity?''

The author of \cite{Kayser}
notes that ''The wave-packet treatment eliminates the need to make some idealizing
assumption by taking both momentum and energy variations properly into account'', and 
illustrates this assertion by comparing his wave-packet 
results with those in the model when the components
$(\nu_1,\nu_2,\nu_3)$ have the same momentum $p$. He introduces the concept
of oscillation lengths $l_{ij}=4\pi p/|\Delta m_{ij}^2|$ where $\Delta m_{ij}^2=m_i^2-m_j^2$.
However, he notes that the wave-packet consideration is model dependent.
We will see below that, for example, in the model when the components have
the same velocity, the oscillation lengths depend not on $m_i^2-m_j^2$ but on 
$m_i-m_j$. Also, the author notes that his consideration is valid on
examples where distances are less than an
extremely large multiple of an oscillation length but is not valid at least for extraterrestrial neutrinos.

In the model when the components
$\nu_j$ are ultrarelativistic and have the same momentum $p$, their energies $E_j$ can be written as
$E_j = (p_j^2+m_j^2)^{1/2} \approx p+m_j^2/2p$. Suppose that all the $\nu_j$ are moving along the $x$ axis. Then the dependence
of the wave function of $\nu_j$ on $t$ and $x$ can be written as 
$\nu_j(t,x)=exp[-i(p^2+m_j^2)^{1/2}t+ipx]\nu_j$ and it follows from Eq. (1) that
\begin{equation}
f_j(t,x)=exp[-i(t-x)p]\sum_{k=1}^3exp(-i\frac{m_k^2}{2p}t)U_{jk}\nu_k\,\, (j=1,2,3)
\end{equation}
As follows from this expression and Eq. (\ref{inverse}), in units $\hbar=c=1$
\begin{equation}
f_j(t,x)=exp[-i(t-x)p]\sum_{k,l=1}^3exp(-i\frac{m_k^2}{2p}t)U_{jk}(U^{-1})_{kl}f_l\,\, (j=1,2,3)
\label{fj}
\end{equation}
If all the $m_k$ are the same and equal $m$, it follows
from Eq. (\ref{fj}) that $f_j(t,x)=exp[-i(t-x)p-i\frac{m^2}{2p}t]f_j$, i.e., $f_j$ is moving as a 
particle with the mass $m$ and neutrino oscillations do not take place. However,
if the masses $m_k$ are different, the oscillations are possible and their quantitative
details depend on the model for $U_{jk}$. 

On the other hand, in the model when the components $\nu_j$ have the same velocity $v$,
instead of Eq. (\ref{fj}) we will have
\begin{equation}
f_j(t,x)=\sum_{k,l=1}^3exp[\frac{-im_k}{(1-v^2)^{1/2}}(t-vx)]U_{jk}(U^{-1})_{kl}f_l\,\, (j=1,2,3)
\label{fjv}
\end{equation}
In that case, the parametrization of the PMNS matrix will be essentially different because the quantities $U_{jk}$ and the oscillation lengths will depend not on the mass-squared differences $\Delta m_{jk}^2$ but on the mass differences $\Delta m_{jk}=m_j-m_k$.

The problem of the quantities $\Delta m_{jk}^2$ has been
discussed in many publications and typical fitting procedures give for those quantities values
of the order of $10^{-4}ev^2$ (see e.g., \cite{Capozzi}). In view of this observation, consider the following problem.

Neutrinos emitted by distant stars fly to Earth for many years. What are typical states of
those neutrinos when they reach the Earth, i.e., to what extent their interaction with the interstellar
medium is important? An analogous problem was widely discussed for photons emitted
by stars. The approach where the interaction of such photons with the interstellar medium
is important was called "tired light". However, after numerous discussions, physicists have concluded that "tired light" does not explain the available data, and that photons 
reaching the Earth practically do not interact with the interstellar medium. Since the interaction 
of neutrinos emitted by stars with the interstellar medium is much weaker than the interaction 
of the photons, a major part of those neutrinos practically
do not interact with the interstellar medium.

Then the following problem arises. When a neutrino with some flavor is created, its wave
function is a direct sum of three wave packets with the masses $(m_1,m_2,m_3)$. What are
typical distances between these wave packets?
The author of \cite{Kayser} notes that they must be much less than the minimum of 
oscillation lengths $l_{ij}$. However, this estimation seems to be necessary but not sufficient.
Indeed, the size of the wave packets of the created neutrino cannot be greater than the
size of the interaction region where the wave packets were created. For example, for the
first reaction of the solar cycle
\begin{equation}
p+p\to d+e^++\nu_e
\label{solarcycle}
\end{equation}
the created particles are created in the region where two protons interact. The size of this region is expected to be of the order of the proton Compton wave length $10^{-15}m$, 
and it is much smaller than typical oscillation lengths.

Consider now how distances between the created wave packets change over time. In standard
quantum theory, the effect of wave packet spreading (WPS) does not take place in momentum
representation but can be important in coordinate representation. 
At the beginning of quantum theory, this effect has been investigated by de Broglie,
Darwin, Schr\"{o}dinger and others. The fact that WPS is inevitable has been treated by several authors as unacceptable and as an indication that standard quantum theory should be modified. 
At the same time, it has not been explicitly shown that numerical results on WPS are 
incompatible with experimental data. For example, as shown by Darwin,  for macroscopic bodies the effect of WPS is extremely small. Probably, it is also believed that in 
experiments on the Earth with atoms 
and elementary particles, spreading does not have enough time to manifest itself although we have not found an explicit statement on this problem in the literature. 

A natural problem arises what happens to photons which can travel from distant objects to Earth even for billions of years. As shown in \cite{book,position}, standard theory predicts that, as a consequence
of WPS, coordinate wave functions  of such photons will have the size of the order of light years or more. Does this contradict observations? We show that it does and the reason 
is that standard position operator is not consistently defined. We argue in \cite{book,position} that the position operator should be modified such that, in the general case, the position and 
momentum representations are not related to each other by the Fourier transform. 
Then we show that, for the modified position operator and ultrarelativistic particles, the effect of WPS does not take place in coordinate representation as well and contradictions with observations do not arise. Therefore, the sizes of the created wave packets do not change over time. 

If  the momenta of the masses $m_i$ are the same, the velocities of those masses are different. Since the masses
are ultrarelativistic, the differences of velocities are $\Delta v_{jk}=\Delta m_{jk}^2/2p^2$ and, if
$p$ is of the order of $Mev/c$ then after a year, the distances between the masses will be 
of the order of $1m$. So, for example, neutrinos coming from Sirius (which is "only" 8 light years
from the Earth) will be superpositions of 
elementary particles the distances between which will be of the order of $8m$,
but a major part of neutrinos reaching the Earth will be quantum superpositions of 
elementary particles the distances
between which can be of the order of kilometers or more. A problem arises whether interactions
of such neutrinos with detectors on Earth can be still described in terms of ($\nu_e,\nu_{\mu},\nu_{\tau}$). As argued in \cite{Kayser}, oscillations will not occur at such conditions.
On the other hand, in the model where the velocities of the masses $m_i$ are the same, the distances between the wave packets will not change over time.
This problem is of great theoretical interest, but its experimental investigation is
problematic. Most of neutrinos detected by neutrino observatories are either solar neutrinos or
neutrinos produced when energetic particles from space crash into Earth’s atmosphere. So, 
it is very difficult to select cases when a detected neutrino came to Earth from a distant star.

We conclude that, although the phenomenon of neutrino oscillations is well confirmed in many experiments, the above discussion shows that 
the theoretical explanation of this phenomenon in the literature is essentially model dependent and is not based on rigorous physical principles. 

\section{Poincare invariance and de Sitter invariance}
\label{AdS}

As already noted, the main goal of the present paper is to investigate whether the
phenomenon of neutrino oscillations can be explained within the de Sitter 
invariant particle theory, but first we consider standard particle theory based on Poincare symmetry where elementary particles
are described by IRs of the Poincare group or its
Lie algebra. The representation operators of the Poincare algebra commute according to the commutation relations 
\begin{eqnarray}
&&[P^{\mu},P^{\nu}]=0,\quad [P^{\mu},M^{\nu\rho}]=-2i(\eta^{\mu\rho}P^{\nu}-
\eta^{\mu\nu}P^{\rho}),\nonumber\\
&&[M^{\mu\nu},M^{\rho\sigma}]=-2i (\eta^{\mu\rho}M^{\nu\sigma}+\eta^{\nu\sigma}M^{\mu\rho}-
\eta^{\mu\sigma}M^{\nu\rho}-\eta^{\nu\rho}M^{\mu\sigma})
\label{PCR}
\end{eqnarray}
where $\mu,\nu=0,1,2,3$, $P^{\mu}$ are the operators of the four-momentum, $M^{\mu\nu}$ are the operators of Lorentz angular momenta and $\eta^{\mu\nu}$ is such that
$\eta^{00}=-\eta^{11}=-\eta^{22}=-\eta^{33}=1$ and $\eta^{\mu\nu}=0$ if 
$\mu \neq \nu$. The coefficient 2 in the r.h.s. of the second and third
expressions is related to the fact that, as explained below, we prefer to work in the
system of units where $\hbar=2$.

Although the Poincare group is the group of motions of Minkowski space, 
the description in terms of relations (\ref{PCR}) does not involve Minkowski space at all.
It involves only representation operators of the Poincare algebra,
and {\it those relations can be treated as a definition of relativistic invariance on
quantum level} (see the discussion in \cite{book,DS}). In particular, the fact that 
$\eta^{\mu\nu}$ formally coincides with the metric tensor in Minkowski space does
not imply that this space is involved.

A background space has a direct physical meaning only on classical level.
Indeed, according to the principles of quantum theory, every physical quantity should be described by an operator but there is no operator corresponding to the point $x$ of
Minkowski space. In quantum theory neither time nor coordinates can be measured with the absolute accuracy (see e.g., a discussion in \cite{book,DS}). In QED, QCD and electroweak theory, the Lagrangian density 
depends on the four-vector $x$ which is associated with a point in Minkowski space but this is only the
integration parameter which is used in the intermediate stage. The goal of the theory is to construct the
$S$-matrix and, when it is already constructed,
one can forget about Minkowski space because no physical quantity depends on $x$. This is in the
spirit of the Heisenberg $S$-matrix program according to which, in relativistic quantum theory it is possible to describe only transitions of states from the infinite past when $t\to -\infty$ to the distant future 
when $t\to +\infty$. 

The fact that covariant local fields play only an auxiliary role for constructing the S-matrix is
also clear from the Pauli theorem \cite{Pauli2} that in the case of fields with an integer spin there is no invariant subspace where the spectrum of the charge operator has a definite sign
while in the case of fields with a half-integer spin there is no invariant subspace where the spectrum of the energy operator has a definite sign. 

Historically, the success of the Dirac equations was a consequence of two major facts:
\begin{itemize}
\item 1) The existence of negative energy solutions of these equations was treated as an
indication of the existence of antiparticles, and shortly after the creation of the Dirac equations, this was confirmed by the discovery of the positron.
\item 2) In the approximation $(v/c)^2$, the Dirac equations describe the fine structure
of the hydrogen atom.
\end{itemize}
As far as item 1) is concerned, we note that, in modern quantum theories, the existence of antiparticles is treated as a 
consequence of the fact that there exist IRs of the Poincare algebra by selfadjoint operators with both, positive and negative energies, and in the Fock space, the energies of all states become positive after the second quantization.

As far as item 2) is concerned, we note that the Dirac equations do not have a rigorous probabilistic interpretation because the representations of the Poincare group describing
transformations of the solutions of these equations are not unitary. However, in the approximation $(v/c)^2$, the probabilistic interpretation is restored,
and in this approximation the Dirac equations describe the fine structure
of the hydrogen atom with a high accuracy because the fine structure constant $\alpha$
is small.

Note that the fact that the $S$-matrix is the operator in momentum space does not exclude a possibility that
in some situations, it is possible to have a space-time description with some accuracy but not with absolute accuracy. First of all, the problem of time is one of
the most important unsolved problems of quantum theory (see e.g., \cite{book} and references therein), and time cannot be measured with the accuracy better than $10^{-18}s$. Also, in typical situations, the position operator in momentum representation exists not only
in the nonrelativistic case but in the relativistic case as well. In the latter case it is known, for example, as
the Newton-Wigner position operator or its modification (see e.g., \cite{position}). As pointed
out even in textbooks on quantum theory, the coordinate description of elementary particles can work only in some approximations. In particular, even in most favorable scenarios, for a massive particle with the mass $m$, its coordinate cannot be measured with the accuracy better than the particle Compton wave length $\hbar/mc$ \cite{BLP}.

Let us also note that in standard particle theories, the origin of masses of elementary
particles in these theories is not discussed. For example, QED does not pose the problem
why the electron mass is as is, and the electroweak theory does not pose the problem
why the neutrino mass is as is. Moreover, the very goal of the renormalization
is to reformulate the theory in such a way that the S-matrix acts in the Fock space where
the elementary particles have their observable masses and charges. In the Standard Model, 
the leptons are originally massless, and their masses arise as a consequence of 
their couplings to the Higgs field. Although the Standard Model has had many successes, it is not yet a complete theory of fundamental interactions, and in the present paper we will not
discuss the origin of masses.

Since $W$ is the Casimir operator of the Poincare algebra,
the mass $m$ of an
elementary particle remains the same during the lifetime of this particle, and, 
since the momentum, energy, and angular momentum operators
commute with the Hamiltonian, 
for any free elementary particle (i.e., the particle which does not interact with other particles and with external fields) these physical quantities are conserved.  

By analogy with relativistic quantum theory, {\it the definition of quantum dS symmetry
should not involve dS space}. If $M^{ab}$ ($a,b=0,1,2,3,4$, $M^{ab}=-M^{ba}$) are the operators describing the system under consideration, then, {\it by definition of dS symmetry on quantum level}, they should satisfy the
commutation relations {\it of the dS Lie algebra} so(1,4), {\it i.e.},
\begin{equation}
[M^{ab},M^{cd}]=-2i (\eta^{ac}M^{bd}+\eta^{bd}M^{ac}-
\eta^{ad}M^{bc}-\eta^{bc}M^{ad})
\label{CR}
\end{equation}
where 
$\eta^{00}=-\eta^{11}=-\eta^{22}=-\eta^{33}=-\eta^{44}=1$ and $\eta^{ab}=0$
if $a\neq b$, and the reason of the coefficient 2 in the r.h.s. of this expression is the same
as in Eq. (\ref{PCR}).
The {\it definition} of AdS symmetry on quantum level is given by the same equations
but $\eta^{44}=1$.

By analogy with Minkowski space, in quantum theories based on dS or AdS symmetry,  dS and AdS spaces have a direct physical meaning only at the classical level, while at the quantum level they can be only auxiliary mathematical concepts in quantum field theory (QFT) for constructing operators satisfying the commutation relations of the dS or AdS algebra. When the theory has already been constructed, one can forget about dS and AdS spaces. 

Let us note that, although QFT has impressive successes in describing experimental data, it also has foundational problems which have been discussed by many authors. One of the main problems in substantiating QFT is that it contains products of interacting quantized fields at the same points. 
As explained in textbooks (e.g., in the book \cite{Bogolubov}), such fields can be treated only as distributions, and the product of distributions at the same point is not a correct mathematical operation. As a consequence, in QFT there are divergences and other inconsistencies. It is rather strange that many physicists believe that such products are needed to preserve locality. However,  as explained above, $x$ is not a physical quantity at quantum level. 

In summary, it does not follow from fundamental principles of quantum theory
that the ultimate quantum theory will be necessarily based on QFT, but a necessary
requirement of dS and AdS quantum theories is that the operators should satisfy the
commutation relations of the dS or AdS algebra, respectively.

The procedure of contraction from dS and AdS symmetries to Poincare one is defined as follows. If we {\it define} the
operators $P^{\mu}$ as $P^{\mu}=M^{4\mu}/2R$, where $R$ is a parameter with the dimension $length$, then in the formal
limit when $R\to\infty$, $M^{4\mu}\to\infty$ but the quantities
$P^{\mu}$ are finite, Eqs. (\ref{CR}) become Eqs. (\ref{PCR}). This procedure is the same for the dS and AdS symmetries. Let us note that, at the quantum level, $R$ is only a parameter of
contraction from the dS and AdS algebras to the Poincare algebra, and, at this level, this parameter has nothing to do with the relation between the de Sitter and Minkowski spaces. 
Let us also note that the result of the famous Dyson's paper \cite{Dyson} that at the quantum level
de Sitter symmetries are more general than Poincare symmetry is based only on the
properties of de Sitter and Poincare groups and also has nothing to do with the relation between the de Sitter and Minkowski spaces. As explained in \cite{book,Symm}, the quantity $R$ can be
treated as the radius of the de Sitter space only in semiclassical approximation, and in this
approximation one obtains for the cosmological acceleration the same result as in
General Relativity which is a pure classical theory.

By analogy with relativistic quantum theory, in theories with dS and AdS symmetries,
elementary particles can be defined as objects described by IRs of the dS and AdS algebras
while the background spaces in those theories play only an auxiliary role and have a direct physical meaning only on classical level. There is a wide literature considering constructions
of local fields in de Sitter invariant theories, e.g., in the framework of AdS/CFT. However,
the ultimate goal of such constructions is to construct the operators $M^{ab}$ describing the system
under consideration in the framework of representation of the de Sitter algebras by
selfadjoint operators in Hilbert spaces. By analogy with standard quantum theory, such
operators act in the Fock space the elements of which are elementary particles
described by IRs of the dS or AdS algebras. By analogy with the above consideration of standard quantum theory,
we will not discuss theories considering the origin of masses in dS or AdS quantum theories
and will assume that in those theories the dS and AdS masses of
elementary particles (see below) should be in agreement with their experimental values.

The  contraction procedure shows that $M^{40}$ is the dS or AdS analog of the relativistic
Hamiltonian and the operators $M^{4i}$ ($i=1,2,3$) are the de Sitter analogs of the
relativistic momentum operators $P^i$. However, as follows from Eq. (\ref{CR}),
the operators $M^{4i}$ do not commute with $M^{40}$. Therefore, if, by analogy with
the relativistic case, the evolution over time is defined by $M^{40}$ then the physical
quantities corresponding to the $M^{4i}$ are not conserved even for free particles.
Also, since the operator ${\tilde W}=M_{04}^2-\sum_{i=1}^3 M_{i4}^2$ is not the Casimir operator
for the dS and AdS algebras, the eigenvalues of this operator cannot be treated
as the mass squared which remains the same during the lifetime of the free elementary
particle. On the other hand, since the operators ($M_{12},M_{31},M_{23}$) commute with
$M^{40}$, the angular momentum operators in de Sitter theories are conserved.

The Casimir operator of the second order for the representation of the so(2,3) algebra is  
\begin{equation}
I_2=\frac{1}{2}\sum_{ab} M_{ab}M^{ab}
\label{I2AdS}
\end{equation}
From now on, the notation $\mu>0$ will be used such that 
states of elementary particles in the AdS invariant theory are eigenvectors of 
$I_2$ with the eigenvalues $\mu^2$, i.e., $\mu$ is the AdS analog of $m$ and
remains the same during the lifetime of the particle.

The procedure of contraction has a physical meaning only if $R$ is rather large. In that case,
$\mu$ and $m$ are related as $\mu=2Rm$, and the relation between the
AdS and Poincare energies is analogous. Since AdS symmetry is more general (fundamental) then Poincare one then 
$\mu$ is more general (fundamental) than $m$. In contrast to the Poincare masses and energies, the AdS masses
and energies are dimensionless. From cosmological considerations (see e.g., \cite{book,Symm}), the value of $R$ is usually accepted to be of the order of $10^{26}m$. Then the AdS masses of the electron,
the Earth and the Sun are of the order of $10^{39}$, $10^{93}$ and $10^{99}$, respectively. 
The fact that even the AdS mass of the electron is so large might be an
indication that the electron is not a true elementary particle.
In addition, the accepted upper level for the photon mass is $10^{-17}ev$. This value
seems to be an extremely tiny quantity. However, the corresponding AdS mass is of the order of $10^{16}$,
 and so, even the mass which is treated as extremely small in Poincare
invariant theory might be very large in AdS invariant theory. 
As noted in Sec. \ref{intr}, it has been proved in \cite{Dyson,book,DS} that quantum theory based on the dS and AdS symmetries is more general (fundamental) than quantum theory based on Poincare symmetry: the latter is the special degenerate case of the former in the
formal limit $R\to\infty$. 

The goal of the present paper can now be formulated as follows. We will not discuss how
the AdS analog of standard electroweak theory is constructed and will consider only
neutrinos created in the framework of this analog. Then, after their creation, the neutrinos
become free particles, i.e., they do not take part in any interactions and are described by IRs of the AdS algebra. A problem arises whether it is possible to explain why, during their flight, those
neutrinos can change their flavors. As noted above, in contrast to standard theory, where
the momentum of a free particle is the same during the whole lifetime of the particle, the AdS
analog of the momentum is constantly changing during the flight and the question arises whether this can be the reason for the change of flavor.

\section{Massless neutrino in AdS theory}
\label{massless}

Before the discovery of neutrino oscillations, the neutrino was treated as the massless
elementary particle with the spin 1/2. In this section we discuss how massless 
elementary particles should be described in AdS theory. Unitary IRs of the AdS group
have been first constructed by Evans \cite{Evans}, unitary IRs of the AdS algebra has been constructed in \cite{JMP}, and the case of massless IRs has been discussed in \cite{tmf}. Below we describe those constructions.

Let $(a_j',a_j'',h_j)$ $(j=1,2)$ be two independent sets of operators satisfying the commutation
relations for the sp(2) algebra \begin{equation}
[h_j,a_j']=-2a_j',\quad [h_j,a_j'']=2a_j'',\quad [a_j',a_j'']=h_j
\label{V9}
\end{equation}
The sets are independent in the sense that for different values of $j$
they mutually commute with each other.

Since the AdS algebra is 10-dimensional, as well as the Poincare algebra, in addition
to the six operators $(a_j',a_j'',h_j)$, we should also define four operators which we denote
as $b', b'',L_+,L_-$. The operators
$L_3=h_1-h_2,L_+,L_-$ satisfy the commutation relations of the su(2) algebra
\begin{equation}
[L_3,L_+]=2L_+,\quad [L_3,L_-]=-2L_-,\quad [L_+,L_-]=L_3
\label{su2CR}
\end{equation}
and the other commutation relations are as follows
\begin{eqnarray}
&[a_1',b']=[a_2',b']=[a_1'',b'']=[a_2'',b'']=[a_1',L_-]=[a_1'',L_+]=\nonumber\\
&[a_2',L_+]=[a_2'',L_-]=0,\quad [h_j,b']=-b',\quad [h_j,b'']=b''\nonumber\\
&[h_1,L_{\pm}]=\pm L_{\pm},\quad [h_2,L_{\pm}]=\mp L_{\pm},\quad [b',b'']=h_1+h_2\nonumber\\
&[b',L_-]=2a_1',\quad [b',L_+]=2a_2',\quad [b'',L_-]=-2a_2'',\quad [b'',L_+]=-2a_1''\nonumber\\
&[a_1',b'']=[b',a_2'']=L_-,\quad [a_2',b'']=[b',a_1'']=L_+ \nonumber\\
&[a_1',L_+]=[a_2',L_-]=b',\quad [a_2'',L_+]=[a_1'',L_-]=-b''
\label{V11}
\end{eqnarray}
At first glance these relations might seem rather chaotic but
in fact they are very natural in the Weyl basis of the so(2,3)
algebra.

The relation between the above set of ten operators and $M_{ab}$ is
\begin{eqnarray}
&M_{10}=i(a_1''-a_1'-a_2''+a_2'),\quad M_{14}=a_2''+a_2'-a_1''-a_1'\nonumber\\
&M_{20}=a_1''+a_2''+a_1'+a_2', \quad M_{24}=i(a_1''+a_2''-a_1'-a_2') \nonumber\\
&M_{12}=L_3,\quad M_{23}=L_++L_-,\quad M_{31}=-i(L_+-L_-)\nonumber\\
&M_{04}=h_1+h_2,\quad M_{34}=b'+b'',\quad M_{30}=-i(b''-b')
\label{V12}
\end{eqnarray}
and therefore the sets are equivalent. 

We work in the system of units $\hbar/2=c=1$. Then $s=1$ for particles with spin 1/2
in the usual units where $\hbar=1$. We use the basis in which the operators $(h_j,K_j)$ 
$(j=1,2)$
are diagonal. Here $K_j$ is the Casimir operator 
\begin{equation}
K_j=h_j^2-2h_j-4a_j''a_j'=h_j^2+2h_j-4a_j'a_j''
\label{V3}
\end{equation}
for the algebra $(a_j',a_j",h_j)$. For constructing IRs we need
operators relating different representations of the
sp(2)$\times$sp(2) algebra. 
By analogy with \cite{book,Evans}, one of the possible choices is:
\begin{eqnarray}
&A^{++}=b''(h_1-1)(h_2-1)-a_1''L_-(h_2-1)-a_2''L_+(h_1-1)
+a_1''a_2''b'\nonumber\\
&A^{+-}=L_+(h_1-1)-a_1''b',\quad
A^{-+}=L_-(h_2-1)-a_2''b',\quad A^{--}=b'
\label{V13}
\end{eqnarray}
We consider the action of these operators only on the space of
minimal sp(2)$\times$sp(2) vectors, i.e., such vectors
$x$ that $a_j'x=0$ for $j=1,2$, and $x$ is the eigenvector of
the operators $h_j$. If $x$ is a minimal vector such that
$h_jx=\alpha_jx$ then $A^{++}x$ is the minimal eigenvector of
the operators $(h_1,h_2)$ with the eigenvalues $(\alpha_1+1,\alpha_2+1)$,
$A^{+-}x$ - with the eigenvalues $(\alpha_1+1,\alpha_2-1)$,
$A^{-+}x$ - with the eigenvalues $(\alpha_1-1,\alpha_2+1)$, and
$A^{--}x$ - with the eigenvalues $(\alpha_1-1,\alpha_2-1)$.

In the theory of IRs of Lie algebras, it is known that each nonzero vector $e_0$ in the space of the IR is cyclic, i.e., that any vector in the representation space can be
obtained by acting by representation operators on $e_0$ and taking all possible 
linear combinations of the results. We choose as $e_0$ the
vector $e_0$ 
satisfying the conditions
\begin{equation}
a_j'e_0=b'e_0=L_+e_0=0,\quad h_je_0=q_je_0\quad (j=1,2)
\label{V15}
\end{equation}
where $q_j$ are positive integers. Then, as shown in \cite{book}, for the massless IR
with the spin $s=1$, $q_2=1$ and $q_1=q_2+s=2$. The reasons why such IRs are called
massless will be explained below.

As follows from Eqs. (\ref{V9}-\ref{V12})
\begin{equation}
I_2=2(h_1^2+h_2^2-2h_1-4h_2-2b''b'+2L_-L_+-4a_1''a_1'-4a_2''a_2')
\label{I2B}
\end{equation}
Then, in the IR characterized by $(q_1,q_2)$,
all the nonzero elements of the representation space are the eigenvectors of the operator $I_2$ with the eigenvalue
\begin{equation}
I_2=2(q_1^2+q_2^2-2q_1-4q_2)
\label{I2C}
\end{equation}
In particular, for the massless neutrino, $I_2=-6$. One of the reasons why such an IR
is called massless, is that, as follows from the contraction procedure, the Casimir operator
$W$ in the Poincare invariant theory is a formal limit of $I_2/4R^2$ when $R\to\infty$
and this limit obviously equals zero.

As follows from Eqs. (\ref{V9}) and (\ref{V11}), the operators
$(a_1',a_2',b')$ reduce the AdS energy $(h_1+h_2)$ by two units. Therefore
$e_0$ is an analog of the state with the minimum energy which can
be called the rest state. In standard classification \cite{Evans}, the
massive case is characterized by the condition $q_2>1$ and the
massless one --- by the condition $q_2=1$.  

We define the vectors
\begin{equation}
e(0,0,1)=[b"(h_1-1)-a_1"L_-]e_0,\,\, f(0,0,0)=L_-e_0,\,\, f(0,0,1)=[b''(h_2-1)-a_2''L_+]f_0
\label{ef}
\end{equation}
Then, as shown in Chapter 8 of \cite{book}, the
basis of the massless IR of the so(2,3) algebra with $s=1$ consists of $e_0$,
the vectors defined by Eq. (\ref{ef}), the vectors
\begin{equation}
e(n_1,n_2,n)=(a_1")^{n_1}(a_2")^{n_2}e(0,0,n),\,\, f(n_1,n_2,n)=(a_1")^{n_1}(a_2")^{n_2}f(0,0,n)
\end{equation} 
if $n<=1$ and the vectors
\begin{eqnarray}
&e(n_1,n_2,n)=(a_1")^{n_1}(a_2")^{n_2}(A^{++})^{n-1}e(0,0,1)\nonumber\\ &f(n_1,n_2,n)=(a_1")^{n_1}(a_2")^{n_2}(A^{++})^{n-1}f(0,0,1)
\end{eqnarray}
if $n>1$. Here $n_1$ and $n_2$ are any positive integers.

One might think that, as follows from the definition of the operators $A^{+-}$ and $A^{-+}$, 
$A^{+-}e(0,0,n)$ should be proportional to $f(0,0,n)$ and 
$A^{-+}f(0,0,n)$ should be proportional to $e(0,0,n)$. However, a direct calculation using 
Eq. (\ref{V11}) shows that, in the massless case, $A^{+-}e(0,0,n)=0$ and $A^{-+}f(0,0,n)=0$.

In Poincare invariant theory without spatial reflections, massless particles
are characterized by the condition that they have a
definite helicity. The operator $M^{04}$ is the AdS energy, and its minimum 
eigenvalue for massless IRs with positive energy is
$E_{min}=3$ for the neutrino. In contrast to the situation in
Poincare invariant theory, where massless particles cannot be
in the rest state, the massless particles in the AdS theory do
have rest states and, for $s=1$, the values of the $z$
projection of the spin can be -1 or 1. However, since $A^{+-}e(0,0,n)=0$ and $A^{-+}f(0,0,n)=0$ for $n\geq 1$, for any value of the energy
greater than $E_{min}$, the spin state is
characterized only by helicity, which can take the values
either 1 or -1, i.e., we have
the same result as in Poincare invariant theory. Note that, in
contrast to IRs of the Poincare algebra, IRs
describing particles in AdS invariant theory belong to the
discrete series of IRs and the energy spectrum in them is
discrete: $E=E_{min}, E_{min}+2, ...\infty$. Therefore,
strictly speaking, the rest states do not have measure zero as in
Poincare invariant theories. 

Nevertheless, although the
probability that the energy is exactly $E_{min}$ is extremely
small, as a consequence of existence of rest states, one IR now contains states with both helicities, 1 and -1. Indeed, the states $e(0, 0, n)$ for $n\geq 1$
have the energy $E=E_{min}+2n$ and helicity 1. By acting by the operator $(b')^n$ on 
$e(0,0,n)$
we obtain the element proportional to the rest state $e_0$. Then $L_-e_0$ is the rest state $f(0,0,0)$.
Finally, as explained above, one can obtain $f(0,0,1)$ from $f(0,0,0)$ and, if $n>1$, $f(0,0,n)$ can be obtained as
$(A^{++})^{n-1}f(0,0,1)$. This state has the energy $E=E_{min}+2n$ and helicity -1. Therefore, as a
consequence of existence of rest states, states with the same energies but opposite helicities belong
to the same IR. 

A known case in Poincare invariant theory is that if neutrino is massless then neutrino and antineutrino are
different particles with opposite helicities. However, in AdS theory, neutrino and antineutrino can be only
different states of the same particle. In experiment they manifest as different particles because
the probability to be in the rest state is extremely small and in weak reactions only states with definite helicities can take part. Also note that, in Poincare invariant theory, the photon
is a massless elementary particle which does not have a definite helicity because it is described by an IR in the theory where, in addition to Poincare invariance, invariance under spatial reflections is required.

\section{Massive spinless elementary particles}
\label{massive}

In what follows, massive elementary particles will be described in semiclassical approximation.
In this approximation, spin effects are not important and therefore we can consider massive
IRs with zero spin. In that case the formulas (\ref{V9}-\ref{V13}) remain valid but,
instead of Eq. (\ref{V15}) we choose as $e_0$ the vector satisfying the conditions
\begin{equation}
a_j'e_0=b'e_0={\bf L}e_0=0,\quad h_je_0=qe_0\quad (j=1,2)
\label{m1}
\end{equation}
where $q>1$.
Since the AdS energy is $(h_1+h_2)$ and the AdS mass is the minimum value of the
AdS energy, then $\mu=2q$. 
Now the elements $e(n_1,n_2,n)$ can be defined as
\begin{equation}
e(n_1,n_2,n)=(a_1")^{n_1}(a_2")^{n_2}(A^{++})^ne_0
\label{m2}
\end{equation}
for all integers $(n_1,n_2,n)\geq 0$.

As follows from Eqs. (\ref{V11}), (\ref{V12}) and (\ref{m1}), the elements $e(n_1,n_2,n)$ are
the eigenvectors of the AdS energy $M_{04}=h_1+h_2$ with the eigenvalues
$\mu+2n+2n_1+2n_2$. Therefore, in the massive case, the AdS energies also take
the values $E=E_{min}, E_{min}+2, ...\infty$ but $E_{min}=\mu$.

\section{Matrix elements of representation operators}
\label{matrix}

The states $e(n_1,n_2,n)$ are not normalized to one. Let 
$N(n_1,n_2,n)=(e(n_1,n_2,n),e(n_1,n_2,n))$ be the norm of the state $e(n_1,n_2,n)$
squared. A direct calculation using Eqs. 
(\ref{V11},\ref{V13},\ref{V15},\ref{ef}) gives that in the case of massless neutrino
\begin{equation}
N(n_1,n_2,n)=n_1!n_2!(1+n+n_1)!(n+n_2)![n!(n-1)!]^2
\end{equation}
while in the massive case the explicit calculation (see Chap. 8 in \cite{book}) gives
\begin{eqnarray}
N(n_1,n_2,n)=n_1!n_2!n!(q+n)_{n_1}(q+n)_{n_2}(\mu-2)_nq_n[(q-1)_n]^3
\end{eqnarray} 
where $q_n=q(q+1)(q+2)\cdots (q+n-1)$ is the Pochhammer symbol.
Therefore, the basis vectors normalized to one are 
${\tilde e}(n_1,n_2,n)=N(n_1,n_2,n)^{-1/2}e(n_1,n_2,n)$. 

As follows from
Eqs. (\ref{V11},\ref{V13},\ref{V15},\ref{ef}), in the massive case, the matrix elements of the representation operators in the normalized basis are given by
\begin{eqnarray}
&b'{\tilde e}(n_1,n_2,n)=[\frac{n_1n_2(\mu+n-2)(n+1)}{(q+n)(q+n-1)}]^{1/2}{\tilde e}(n_1-1,n_2-1,n+1)+\nonumber\\
&[\frac{n(\mu+n-3)(q+n+n_1-1)(q+n+n_2-1)}{(q+n-2)(q+n-1)}]^{1/2}{\tilde e}(n_1,n_2,n-1)\nonumber\\
&b"{\tilde e}(n_1,n_2,n)=[\frac{(n+1)(q+n+n_1)(q+n+n_2)(\mu+n-2)}{(q+n)(q+n-1)}]^{1/2}{\tilde e}(n_1,n_2,n+1)+\nonumber\\
&[\frac{n(\mu+n-3)(n_1+1)(n_2+1)}{(q+n-1)(q+n-2)}]^{1/2}{\tilde e}(n_1+1,n_2+1,n-1)\nonumber\\
&a_1'{\tilde e}(n_1,n_2,n)=[n_1(q+n+n_1-1)]^{1/2}{\tilde e}(n_1-1,n_2,n)\nonumber\\ 
&a_1"{\tilde e}(n_1,n_2,n)=[(n_1+1)(q+n+n_1)]^{1/2}{\tilde e}(n_1+1,n_2,n)
\nonumber\\
&a_2'{\tilde e}(n_1,n_2,n)=[n_2(q+n+n_2-1)]^{1/2}{\tilde e}(n_1,n_2-1,n)\nonumber\\ 
&a_2"{\tilde e}(n_1,n_2,n)=[(n_2+1)(q+n+n_2)]^{1/2}{\tilde e}(n_1,n_2+1,n)
\nonumber\\
&L_+{\tilde e}(n_1,n_2,n)=[\frac{n_2(n+1)(q+n+n_1)(\mu+n-2)}{(q+n)(q+n-1)}]^{1/2}{\tilde e}(n_1,n_2-1,n+1)+\nonumber\\
&[\frac{n(\mu+n-3)(n_1+1)(q+n+n_2-1)}{(q+n-1)(q+n-2)}]^{1/2}{\tilde e}(n_1+1,n_2,n-1)\nonumber\\
&L_-{\tilde e}(n_1,n_2,n)=[\frac{n_1(n+1)(q+n+n_2)(\mu+n-2)}{(q+n)(q+n-1)}]^{1/2}{\tilde e}(n_1-1,n_2,n+1)+\nonumber\\
&[\frac{n(\mu+n-3)(q+n+n_1-1)(n_2+1)}{(q+n-1)(q+n-2)}]^{1/2}{\tilde e}(n_1,n_2+1,n-1)\nonumber\\
&h_1{\tilde e}(n_1,n_2,n)=(q+n+2n_1){\tilde e}(n_1,n_2,n),\,\, 
h_2{\tilde e}(n_1,n_2,n)=(q+n+2n_2){\tilde e}(n_1,n_2,n)
\label{matrixelements}
\end{eqnarray}

For massless neutrino, we will consider only the case $(n_1,n_2,n)\gg 1$. Then the
matrix elements are defined by the same expressions where $\mu=2,\,\,q=1$.

At the present stage of the universe, the value of $R$ is very large. Therefore, 
in situations where Poincare limit is valid with a high accuracy, the quantum numbers 
$(n_1, n_2, n)$ are very large since in the formal limit $R\to\infty$  the quantities 
$(n_1/2R, n_2/2R, n/2R)$ become continuous momentum variables.
For this reason, the wave functions of the neutrino and spinless massive particles 
can be written as
\begin{equation}
\Psi=\sum_{n_1n_2n} c(n_1,n_2,n){\tilde e}(n_1,n_2,n)
\label{Psi}
\end{equation}
where the minimum value of $n$ is very large. 

Since the current theory of weak interactions is based on Poincare
invariance, the form of the neutrino and lepton wave functions in Eq. (\ref{Psi}) is a problem. In standard
particle theory, particles in Feynman diagrams have definite four-momenta
and such state states are not even normalized. When it is said that a
particle has the momentum ${\bf p}_0$, it is assumed that 
the state of the particle is described by a wave function
$\int c({\bf p})|{\bf p}>d^3{\bf p}$
where $c({\bf p})$ has a sharp maximum at ${\bf p}_0$ and the width of the maximum is much less
than $|{\bf p}_0|$. As noted above, in the approach proposed in \cite{Naumov} the neutrinos are
described by wave packets.

However, in the AdS case, the situation is more complicated. The AdS analogs of the momentum
operators are $M^{4i}\,\,(i=1,2,3)$ and these operators do not commute with each other. In addition, IRs of the AdS algebra belong to the discrete series. Therefore, it is a problem
how to generalize Feynman diagrams to the AdS case. Nevertheless, when the quantum numbers 
$(n_1, n_2, n)$ are very large, one might
expect that semiclassical approximation will work with a high accuracy because,
as follows from Eqs. (\ref{matrixelements}), the operators of the IR of the AdS
algebra can change these numbers only by one.

\section{Semiclassical approximation}
\label{semicl}

A typical form of the semiclassical wave function is
\begin{equation}
c(n_1, n_2, n) = a(n_1, n_2, n)exp[i (n_1\varphi_1 + n_2\varphi_2+n\varphi)]
\label{c(n1n2n)}
\end{equation}
where the amplitude $a(n_1, n_2, n)$ has a sharp maximum at semiclassical values of
$(n_1, n_2, n)$. Since these numbers are very large, when some of them change
by one, the major change of $c(n_1, n_2, n)$ comes from the rapidly oscillating exponent
and in semiclassical approximation the change of $a(n_1, n_2, n)$ can be neglected.
As a consequence, in semiclassical approximation, each representation operator
becomes the operator of multiplication by a function and, as follows from Eqs.
(\ref{V12}) and (\ref{matrixelements}), if $\varphi_1=\pi+\chi_1$ then 
\begin{eqnarray}
&M_{14}=2[n_1(q+n+n_1)]^{1/2}cos\chi_1+
2[n_2(q+n+n_2)]^{1/2}cos\varphi_2\nonumber\\
&M_{24}=-2[n_1(q+n+n_1)]^{1/2}sin\chi_1+2[n_2(q+n+n_2)]^{1/2}sin\varphi_2\nonumber\\
&M_{34}=\frac{2[n(\mu+n)]^{1/2}}{q+n}\{[(q+n+n_1)(q+n+n_2)]^{1/2}cos\varphi-
(n_1n_2)^{1/2}cos(\chi_1+\varphi_2-\varphi)\}\nonumber\\
&M_{10}=-2[n_1(q+n+n_1)]^{1/2}sin\chi_1-2[n_2(q+n+n_2)]^{1/2}sin\varphi_2\nonumber\\
&M_{20}=-2[n_1(q+n+n_1)]^{1/2}cos\chi_1+2[n_2(q+n+n_2)]^{1/2}cos\varphi_2\nonumber\\
&M_{30}=\frac{2[n(\mu+n)]^{1/2}}{q+n}\{-[(q+n+n_1)(q+n+n_2)]^{1/2}sin\varphi+
(n_1n_2)^{1/2}sin(\chi_1+\varphi_2-\varphi)\}\nonumber\\
&M_{31}=\frac{2[n(\mu+n)]^{1/2}}{q+n}\{[n_1(q+n+n_2)]^{1/2}sin(\chi_1-\varphi)
+[n_2(q+n+n_1)]^{1/2}sin(\varphi_2 -\varphi)\}\nonumber\\
&M_{23}=\frac{2[n(\mu+n)]^{1/2}}{q+n}\{-[n_1(q+n+n_2)]^{1/2}cos(\chi_1-\varphi)+
[n_2(q+n+n_1)]^{1/2}cos(\varphi_2-\varphi)\}\nonumber\\
&M_{04}=\mu+2(n+n_1+n_2),\quad M_{12}=2(n_1-n_2)
\label{Mab}
\end{eqnarray}

\begin{eqnarray}
&{\tilde W}=(\mu+2n+2n_1+2n_2)^2-4n_1(q+n+n_1)-
4n_2(q+n+n_2)\nonumber\\
&-8[n_1n_2(q+n+n_1)(q+n+n_2)]^{1/2}cos(\chi_1+\varphi_2)\nonumber\\
&-\frac{4n(\mu+n)}{(q+n)^2}\{(n_1n_2)^{1/2}
cos(\chi_1+\varphi_2-\varphi)-[(q+n+n_1)(q+n+n_2)]^{1/2}cos\varphi\}^2\nonumber\\
&{\bf N}^2=M_{10}^2+M_{20}^2+M_{30}^2=\nonumber\\
&4n_1(q+n+n_1)+4n_2(q+n+n_2)-8[n_1n_2(q+n+n_1)(q+n+n_2)]^{1/2}
cos(\chi_1+\varphi_2)+\nonumber\\
&4\frac{n(\mu+n)}{(q+n)^2}
\{[(q+n+n_1)(q+n+n_2)]^{1/2}sin\varphi-(n_1n_2)^{1/2}
sin(\chi_1+\varphi_2-\varphi)\}^2\nonumber\\
&{\bf M}^2=M_{12}^2+M_{31}^2+M_{23}^2=\nonumber\\
&4(n_1-n_2)^2+\frac{4n(\mu+n)}{(q+n)^2}\{
n_1(q+n+n_2)+n_2(q+n+n_1)\nonumber\\
&-2[n_1n_2(q+n+n_1)(q+n+n_2)]^{1/2}cos(\chi_1+\varphi_2-2\varphi)\}
\label{Mab2}
\end{eqnarray}
and, in semiclassical approximation, the results for the neutrino are given by the
same expressions when $\mu=0$.
Since $I_2={\tilde W}-{\bf N}^2+{\bf M}^2$, it follows from Eq. (\ref{Mab2})
that for the neutrino, $I_2=0$ and in the massive case  $I_2=\mu^2$. This is in agreement 
with the exact result for neutrino $I_2=-6$ obtained from Eq. (\ref{I2C})
because in semiclassical approximation, the numbers $(n_1,n_2,n)$
are much greater than 1. 

We will now check whether the above results work in Poincare approximation.
At the present stage of the universe, the value of $R$ is very large and therefore this
approximation should work with a very high accuracy.  As noted in Sec. \ref{AdS}, the
four-momentum $P_{\nu}= M_{\nu 4}/2R$ should be finite in the formal limit $R\to\infty$
and then, as follows from Eq. (\ref{Mab}), the numbers $(n_1,n_2,n)$ should be very large
because $(n_1/R,n_2/R,n/R)$ should be finite in this limit.
At the same time, even when $R$ is very large, the Lorenz algebra operators $M_{ab}$ 
($a,b\neq 4$) should be finite because they should be the same as in Poincare
approximation. Then, as follows from the expression for $M_{12}$ in Eq. (\ref{Mab}),
the numbers $n_1$ and $n_2$ should be close to each other because $n_1-n_2$
should be finite in the formal limit $R\to\infty$ and, as follows from other expressions
for $M_{ab}$, the angles $(\chi_1,\varphi_2,\varphi)$ should be very small because
they should be of the order of $O(1/R)$.

Therefore, as follows from Eq. (\ref{Mab}), with our choice of the angles $(\chi_1,\varphi_2,\varphi)$, in Poincare approximation, $M_{24}=M_{20}=M_{12}=M_{23}=0$, i.e., the
particle is moving in the $xz$ plane. As follows from Eq. (\ref{Mab}), in this approximation, 
the $x$ and $z$ 
components of the particle momentum are given by
\begin{equation}
p_x=M_{14}/(2R)=\frac{2}{R}[n_1(q+n+n_1)]^{1/2},\quad p_z=M_{34}/(2R)=
\frac{1}{R}[n(\mu+n)]^{1/2}
\label{pxz}
\end{equation}

Since the operator ${\bf N}$ should be the same as in Poincare
approximation, we can use the known result for IRs of the Poincare algebra that, 
in this approximation, this operator in momentum representation has the form (see e.g.,
Chap. 2 in \cite{book}) ${\bf N}=-i\hbar \epsilon({\bf p}) \partial /\partial {\bf p}$ where
$ \epsilon({\bf p})=(m^2+{\bf p}^2)^{1/2}$ is the standard energy in relativistic theory.
In semiclassical approximation, the form of the position operator is described
even in textbooks: ${\bf r}= i\hbar \partial /\partial {\bf p}$, and the commutator of this 
operator with $\epsilon({\bf p})$ can be neglected. Therefore we get that 
${\bf N}=-\epsilon({\bf p}){\bf r}$ and then, as follows from Eq. (\ref{Mab}),
the $x$ and $z$ coordinates of the particle are given by
\begin{equation}
x=-\frac{2}{\epsilon({\bf p})}[n_1(q+n+n_1)]^{1/2}(\chi_1+\varphi_2),\,\,
z=\frac{2[n(\mu+n)]^{1/2}}{\epsilon({\bf p})(q+n)}[-(q+n+2n_1)\varphi +n_1(\chi_1+\varphi_2)]
\label{xz}
\end{equation}

Finally, since in Poincare semiclassical approximation, $M_{31}=L_y=zp_x-xp_z$, and
$q+n+2n_1\approx (\mu+2n+2n_1+2n_2)/2=R\epsilon({\bf p})$, it follows
from Eqs. (\ref{pxz}) and (\ref{xz}) that
\begin{equation}
M_{31}=\frac{2}{q+n}[nn_1(\mu+n)(q+n+n_1)]^{1/2}(\chi_1+\varphi_2-2\varphi)
\end{equation}
and, in the discussed approximation, this result coincides with $M_{31}$ from Eq. (\ref{Mab}).

\section{Neutrino flavors in AdS theory}
\label{flavors}

We now take into account the terms quadratic in $(\chi_1,\varphi_2,\varphi)$ and the 
fact $n_1-n_2$ is of the order of $O(1/R)$. Then the
expression for ${\tilde W}$ in the first expression in Eq. (\ref{Mab2}) becomes: 
\begin{equation}
{\tilde W}=\mu^2+\frac{4n_1(\chi_1+\varphi_2)}{q+n}[(q^2+n_1q+n_1n)
(\chi_1+\varphi_2)+2n(\mu+n)\varphi]+4n(\mu+n)\varphi^2
\label{tildeW}
\end{equation}
In Poincare approximation, i.e., in the formal limit $R\to\infty$, ${\tilde W}/4R^2$ should become
$m^2$. Since $\mu=2mR$ and the angles $(\chi_1,\varphi_2,\varphi)$ are of the order of $O(1/R)$, this is the case. In particular, since we treat neutrino as a particle which in
Poincare approximation becomes massless, then ${\tilde W}/4R^2$ should become zero in
the formal limit $R\to\infty$. As follows from Eq. (\ref{tildeW}), this is the case
because $\mu=0$ for the neutrino and then 
\begin{equation}
{\tilde W}=4(n_1\chi_1+n_2\varphi_2+n\varphi)^2
\label{M2}
\end{equation}

In Poincare theory, the evolution is defined by the operator $exp(-iHt/\hbar)$ where $H$ is
the Hamiltonian and $t$ is time. Since $M_{04}$ is the AdS analog of the Poincare Hamiltonian, and
$M_{04}/2R$ should become $H$ when $R$ is very large, the evolution in AdS
theory is defined by $exp(-iM_{04}/(\tau\hbar)$ where $\tau = t/2R$. Then, as follows from 
Eq. (\ref{c(n1n2n)}) and the first expression in Eq. (\ref{Mab}),  
the evolution of the angles $(\chi_1,\varphi_2,\varphi)$ is defined by
\begin{equation}
\chi_1(t)=\chi_{10}-t/(2R),\quad \varphi_2(t)=\varphi_{20}-t/(2R)\quad 
\varphi(t)=\varphi_0-t/(2R) 
\end{equation}
and then, since $\hbar=2$, it follows from Eq. (\ref{M2}) that
$$ {\tilde W}=4(n_1\chi_{10}+n_2\varphi_{20}+n\varphi_0-Et/2)^2$$
because, as follows from the first expression in Eq. (\ref{Mab}), the Poincare energy $E$ equals 
$(n_1+n_2+n)/R$. Experimental data on neutrino oscillations are usually described in terms
of the neutrino energy and the oscillation distance $l=ct$. Hence, inserting standard values
of $\hbar$ and $c$,
we finally get
\begin{equation}
{\tilde W}=4(A-\frac{El}{2\hbar c})^2
\label{final}
\end{equation}
where 
\begin{equation}
A=A(n_1,n_2,n,\chi_{10},\varphi_{20},\varphi_0)=n_1\chi_{10}+n_2\varphi_{20}+n\varphi_0
\label{A}
\end{equation}

As noted in Sec. \ref{AdS}, in the famous Dyson's paper \cite{Dyson} and in our publications
\cite{book,DS} it has been proved that quantum theory based on dS and AdS symmetries is more
general (fundamental) than quantum theory based on Poincare symmetry.
We believe that it is rather strange that, although the paper \cite{Dyson} appeared more than 50 years ago, standard particle theories (QED, QCD and electroweak theory)
are still based on Poincare symmetry, and possible reasons for such a situation have been
mentioned in Sec. \ref{AdS}.

In Poincare invariant theory, the representation operators $M^{\mu\nu}$ 
($\mu,\nu=0,1,2,3$) of the Lorentz algebra are dimensionless (in units $\hbar=c=1$) while
the momentum operators $P^{\mu}$ have the dimension $1/length$. As noted in Sec.
\ref{AdS}, the AdS invariant theory also contains the operators $M^{\mu\nu}$ but,
instead of the four operators $P^{\mu}$, it contains the dimensionless operators $M^{4\mu}$.
Poincare invariant theory is a special degenerate case of AdS invariant theory such that the operators $P^{\mu}$ can be treated
as a formal limit of $M^{4\mu}/2R$ when $R\to\infty$.

In Poincare invariant theory, the operator $W=(P^0)^2-{\bf P}^2$ is the Casimir operator
and elementary particles are eigenstates of $W$ with eigenvalues $m^2$ where $m$ is called
the mass. In this theory, the energy, the momentum 
and the mass of a free particle do not change during the whole lifetime of the particle.

In AdS theory, there are two analogs of $W$ --- the Casimir operator $I_2$ defined
by Eq. (\ref{I2AdS}) and the operator ${\tilde W}$ defined in Sec. \ref{AdS}. States of elementary
particles are eigenvalues of $I_2$ which remain the same
over the whole lifetime of the particle but {\it even for a free particle, the spectrum of
${\tilde W}$ changes over time}. 

The eigenvalues of ${\tilde W}$ are dimensionless and, in Poincare approximation, they are related
to the eigenvalues of $W$ as $W={\tilde W}/4R^2$ when $R$ is very large. As shown in
Sec. \ref{massless}, in semiclassical approximation, the neutrino is a state with the
eigenvalue of ${\tilde W}$ given by Eq. (\ref{M2}), and this eigenvalue changes over
time. As follows from Eq. (\ref{final}), in general, the eigenvalues ${\tilde w}$ of
${\tilde W}$ are rather large numbers but, when $R\to\infty$, the formal limit of 
${\tilde w}/4R^2$ equals zero, in agreement with treating the neutrino as the massless
particle in Poincare approximation.

The phenomenon of neutrino oscillations shows that, when free neutrinos 
 fly a long distance, they can change their flavor. 
As noted above, in the current treatment of neutrino oscillations, such neutrinos are treated
as quantum superpositions of elementary particles with different masses $(m_1,m_2,m_3)$ and,
depending on the coefficients of the superpositions and the neutrino helicities, 
the neutrinos are treated as one of the
particles  $(\nu_e, \nu_{\mu},\nu_{\tau})$ or their antiparticles. As noted in Sec. \ref{dirsum},
such a treatment is highly problematic from the theoretical point view.

The goal of the present paper is to investigate whether AdS quantum theory can shed a new light
on the neutrino oscillation phenomenon. We treat the neutrino not as
a superposition of three elementary particles but as a particle which in AdS quantum theory
is treated as elementary and massless, and the meaning of such a treatment is explained
in Sec. \ref{massless}. 

As follows from Eq. (\ref{M2}), for neutrinos, eigenvalues of
${\tilde W}$ are rather large but a formal limit of the operator
${\tilde W}/4R^2$ when $R\to\infty$ is zero, in agreement with the fact that this limit equals
$m^2$ and, in our approach, the Poincare neutrino mass equals zero. It will be clear from
our discussion that the neutrino is a quantum superposition of states with essentially different 
eigenvalues ${\tilde w}$ of the operator ${\tilde W}$. {\it Our conjecture is that the neutrino flavor is defined by the most probable range of ${\tilde w}$ in the neutrino wave function.} As noted in Sec. \ref{matrix},
computing the neutrino wave function in the AdS theory is a problem and so, our conjecture will be
proved or disproved when this theory is constructed.

However, for the electron,
muon and tau lepton, the situation is drastically different. For example, for
the electron, typical eigenvalues of ${\tilde W}$ are of the order of $4m_e^2R^2$ where $m_e$ is
the electron mass. As follows from cosmological data (see e.g.,
the discussion in \cite{book}), the value of $R$ is of the order of $10^{26}m$. Then, as follows
from Eq. (\ref{tildeW}), all the eigenvalues
of the operator ${\tilde W}$ are very close to $4m_e^2R^2 \sim 10^{77}$ while for the muon and tau lepton,
almost all the eigenvalues are close to $4m_{\mu}^2R^2 \sim 10^{81}$ and
$4m_{\tau}^2R^2 \sim 10^{84}$, respectively. {\it Therefore, our conjecture naturally explains
the fact that the electron, muon and tau lepton do not have flavors changing over time}.  

The author of \cite{Klute} summarizes the present experimental status
of neutrino oscillations as follows:
\begin{itemize}
\item Atmospheric $\nu_{\mu}$ and $\bar{\nu}_{\mu}$ disappear most likely converting to
$\nu_e$ and $\bar{\nu}_e$. The results show an energy and distance dependence perfectly
described by mass-induced oscillations.
\item Accelerator $\nu_{\mu}$ and $\bar {\nu}_{\mu}$ disappear over distances 
of $\sim$ 200 to 800 {\it km}. The energy spectrum of the results show a clear oscillatory behavior
also in accordance with mass-induced oscillations with wavelength in agreement with the
effect observed in atmospheric neutrinos.
\item Accelerator $\nu_{\mu}$ and $\bar {\nu}_{\mu}$ appear as $\nu_e$ and $\bar{\nu}_e$ 
at distances $\sim$ 200 to 800 {\it km}. 
\item Solar $\nu_e$ convert to $\nu_{\mu}$ and/or $\nu_{\tau}$. The observed energy dependence
of the effect is well described by massive neutrino conversion in the Sun matter according to the MSW effect.
\item Reactor $\bar{\nu}_e$ disappear over distances of $\sim$ 200{\it km} and $\sim$ 1.5{\it km} with 
different probabilities. The observed energy spectra show two different mass-induced oscillation wavelengths:
at short distances in agreement with the one observed in accelerator $\nu_{\mu}$ disappearance,
and a long distance compatible with the required parameters for MSW conversion in the Sun. 
\end{itemize}
Here, for example, the word ''disappear'' means not that all neutrinos in the given experiment
disappear but only a part of them disappears, and the words ''appear" and ''convert'' should be understood
analogously. We will describe the results of experiments where the energies and flavors of initial
and final neutrinos are known without model assumptions.

In the K2K experiment \cite{K2K}, 56 $\nu_{\mu}$ neutrinos with the energies 
$E\approx 1.3Gev$ have been observed at the Super-Kamiokande far detector at 250 km distance while the expectation 
was $80.6_{-8.0}^{+7.3}$. For this case, $El/({2\hbar}c)=0.8\cdot 10^{18}$.

In the KamLAND experiment \cite{KamLAND}, ${\bar \nu}_e$ reactor neutrinos with the energies in the range
$(2.6-6)Mev$ flew the distance 180km and, although 365 events were predicted, only 258 were observed.
For this case, if $E=3Mev$ then $El/({2\hbar}c)\approx 10^{15}$. 

In the RENO experiment \cite{RENO}, ${\bar \nu}_e$ reactor neutrinos with the energies $E\approx 3Mev$
flew the distance 1183m. The ratio of observed to expected numbers of antineutrinos was
$0.920 \pm 0.009(stat.)  \pm 0.014(syst.)$. For this case, $El/(2{\hbar}c)=10^{13}$.

In the NOvA experiment \cite{NOvA}, $\nu_{\mu}$ neutrinos with the energies in the range $(2-20)Gev$
flew from Fermilab 810km. 720 cases were expected without oscillations but only 126 $\nu_{\mu}$
cases and 66 $\nu_e$ cases were observed. For this range of energies, $El/(2{\hbar}c)$ is in the range
$(0.4-4)\cdot 10^{19}$.

In the OPERA experiment \cite{OPERA}, $\nu_{\mu}$ neutrinos with the energies $E\approx 17Gev$ flew from CERN toward the Gran Sasso underground laboratory, 730 km away. 5603 neutrino interactions were fully reconstructed, 10 of them were identified as conversions to $\nu_{\tau}$ and 35 --- as conversions to $\nu_e$. In this experiment, 
$El/(2{\hbar}c)\approx 3.1\cdot 10^{19}$. 

Those experiments confirm that the phenomenon of neutrino oscillation does take place. At the same time, in all those experiment except \cite{NOvA}, most neutrinos did not change their flavor. In the framework of the approach described in Sec. \ref{matrix},
the description of the above experiments is model dependent because, in semiclassical approximation, the values of the angles
$(\chi_{10},\varphi_{20},\varphi_0)$ in Eq. (\ref{A}) are not known. We noted
that those values should be very small. Since in the above experiments, the phenomenon of neutrino oscillations does take place,
it follows from Eq. (\ref{final}) and from the above values of the quantity $El/(2{\hbar}c)$ that ${\tilde W}$ essentially depends on $l$. Therefore the quantity $A$ is of the order of $10^{19}$ or less. Since the AdS neutrino energy equals
$2(n_1+n_2+n)$ then, even if Poincare energy is of the order of $Mev$, the AdS energy is of the order of
$2R\cdot Mev\sim 10^{39}$, i.e., the quantities $(n_1,n_2,n)$ in $A$ are of the order of $10^{39}$ and 
the angles $(\chi_{10},\varphi_{20},\varphi_0)$ are indeed very small.

As noted in Sec. \ref{dirsum}, the current theory of neutrino oscillations cannot describe the data 
without fitting parameters in the
Pontecorvo–Maki–Nakagawa–Sakata matrix. Analogously, as noted in Sec. \ref{matrix}, in the
absence of AdS theory of weak interactions, the details of the
initial neutrino wave function are not known. We propose to consider this function in
semiclassical approximation. Then one can treat $(n_1,n_2,n,\chi_{10},\varphi_{20},\varphi_0)$
as fitting parameters which can be found from experimental values of the quantities
$M^{\mu\nu}$ by using Eq. (\ref{Mab}) but these quantities are not measured in neutrino
experiments. 

However, there is a case which can be described without
fitting parameters. Historically, the problem of neutrino oscillations first appeared in view of the solar neutrino problem. 
In 2002, Ray Davis and Masatoshi Koshiba won part of the Nobel Prize in Physics for experimental work which found the number of solar neutrinos to be around a third of the number predicted by the standard solar model.

The result \cite{solar} of the Sudbury Neutrino Observatory is that the $\nu_e$ survival
probability at 10$Mev$ is $0.317\pm 0.016(stat)\pm 0.009(syst)$, and this number is very close to 1/3. If we take for $l$ the distance from Sun to Earth then for $E=10Mev$, $El/(2{\hbar}c)=3.75\cdot 10^{21}$.
Since the quantity $A$ is of the order of $10^{19}$ or less, one can neglect $A$ in Eq. (\ref{final})
and get 
\begin{equation}
{\tilde W}=(\frac{El}{\hbar c})^2
\label{noA}
\end{equation}
Since most neutrinos created in the Sun are created in the reaction (\ref{solarcycle})
and their typical energies are in the range $(1-20)Mev$, the values of ${\tilde W}$ are
in the range $(0.5\cdot 10^{42}-2.5\cdot 10^{44})$. These values should not be treated as anomalously large. Indeed, as noted above, even for the electron, which is the lightest massive particle, the AdS mass squared is $4m_e^2R^2$ and this quantity is of the order of $10^{77}$.

As noted above, {\it our conjecture is that the neutrino flavor is defined by the most probable range of ${\tilde w}$ in the neutrino wave function.} So, if we assume that the range $(0.5\cdot 10^{42}-2.5\cdot 10^{44})$
is partitioned into equal pieces such that each piece corresponds to one of the neutrino flavors, 
then the $\nu_e$ neutrinos created on the Sun can be detected
on the Earth with equal probabilities 1/3 as $\nu_e$, $\nu_{\mu}$ or $\nu_{\tau}$. So, our
conjecture naturally explains that the survival probability for $\nu_e$ equals 1/3.

As noted in Sec. \ref{dirsum}, the application of the present theory of neutrino oscillations to
neutrinos coming to Earth from distant stars is problematic. On the other hand, in our
approach one can neglect $A$ in Eq. (\ref{final}) with a high accuracy. For example, for
neutrinos coming from Sirius, the values of ${\tilde W}$ are
in the range $(1.5\cdot 10^{53}-7.5\cdot 10^{55})$. However, as already noted, 
detection of such neutrinos  at the present status of experimental technique is not
possible.  

\section{Conclusion}
\label{Conclusion}

One of the main problems in generalizing the standard theory of neutrino interactions to the case of de Sitter invariance is, apparently, the following. In standard theory, the processes of
neutrino absorption/creation are described by Feynman diagrams containing $(W,lepton,\nu)$
vertices where the particles in these vertices have definite four-momenta. The de Sitter
analog of the four-momentum operator $P^{\mu}\,\,(\mu=0,1,2,3)$ is the operator
$M^{4\mu}$. Since different components of the $M^{4\mu}$ do not commute with each
other, there are no de Sitter diagrams where the particles in such vertices
have definite values of $M^{4\mu}$ for all $\mu$. We assume that the future
AdS invariant quantum theory of weak interactions will define possible neutrino
wave functions in the $(W,lepton,\nu)$ vertices. The fact that this theory is more general (fundamental) than
standard Poincare invariant theory, has been proved in the famous Dyson's paper
"Missed Opportunities" \cite{Dyson} and in our publications \cite{book,DS}. Those results have been proved proceeding from the properties of de Sitter and Poincare groups and 
from the properties of de Sitter and Poincare Lie algebras, respectively. The derivation of the results does not involve de Sitter and Minkowski spaces in any way. We believe that it is rather strange that, although the Dyson's paper has appeared more than 50 years ago, standard particle theories
(QED, QCD and electroweak theory) are still based on Poincare symmetry, and in Sec. \ref{intr} we describe possible reasons for such a situation.

In the current approach to neutrino oscillations, the neutrino is treated not as an
elementary particle but as a superposition of three elementary particles with different masses. In Sec. \ref{dirsum} we note that this approach is essentially model dependent and is not based on rigorous physical principles. We propose an approach where the neutrino is treated as a massless elementary particle in AdS
invariant quantum theory. 

If $P_0$ and ${\bf P}$ are the operators of standard energy and momentum then,  
in standard theory, elementary particles are eigenstates of $W=P_0^2-{\bf P}^2$ with the eigenvalue $w=m^2$ where $m$ is called the mass. In this theory, the value of $m$ remains the same
during the whole lifetime of the particle. The operator ${\tilde W}$ defined in Sec. \ref{AdS} is
an AdS analog of $W$. In contrast to the situation in Poincare invariant theory, the spectrum of ${\tilde W}$ changes over time even for elementary particles. If the spectrum of ${\tilde W}$ consists of the values ${\tilde w}$ then {\it our conjecture is that the future AdS theory will show that the neutrino flavor is defined by the most probable range of ${\tilde w}$ in the neutrino wave function.}

In this approach, it becomes clear why, in contrast to the neutrino, the electron, muon and tau lepton do not have flavors changing over time. Since the current theory of weak interactions is based on Poincare
invariance, the calculation of the most probable range of ${\tilde w}$ in neutrino wave functions
is a problem but, as explained in Sec. \ref{flavors}, the solar neutrino problem has a natural explanation. As noted in Secs. \ref{dirsum} and \ref{flavors}, the experimental investigation of neutrinos
coming to Earth from distant stars would be very important for understanding the
mechanism of neutrino oscillations, but, at the present status of experimental technique, such an
investigation is not possible. 

{\bf Acknowledgements:} I am grateful to Norma Susana Mankoč Borštnik for discussions which stimulated me to write this paper. I am also grateful to (anonymous) reviewers of this paper
for important remarks and for pointing my attention to paper \cite{Kayser}. This stimulated me to significantly expand Sections 2 and 3.

\end{document}